\documentclass[showpacs,twocolumn]{revtex4}

\usepackage{amsmath,amssymb,mathrsfs}
\usepackage{latexsym}
\usepackage{graphicx} 

\newcommand{\bs}[1]{\boldsymbol{#1}}
\newcommand{\comm}[2]{\left[#1,#2\right]}

\newcommand{\vac}{\left|\,0\,\right\rangle}
\newcommand{\ket}[1]{\left|#1\right\rangle}

\def\ie{\emph{i.e.},\ }
\def\eg{\emph{e.g.}\ }
\def\ea{\emph{et al.}}

\def\b{\text{b}}
\def\r{\text{r}}
\def\g{\text{g}}

\allowdisplaybreaks[1]
\begin{document}
 \title{Exact results for SU($\bs{3}$) spin chains:\\ 
 trimer states, valence bond solids, and their parent Hamiltonians}
 \author{Martin Greiter, Stephan Rachel, and Dirk Schuricht}
 \affiliation{Institut f\"ur Theorie der Kondensierten Materie,
   Universit\"at Karlsruhe, Postfach 6980, 76128 Karlsruhe, Germany}
 \pagestyle{plain}
 \begin{abstract}
   We introduce several exact models for SU(3) spin chains: (1) a
   translationally invariant parent Hamiltonian involving four-site
   interactions for the trimer chain, with a three-fold degenerate
   ground state.  We provide numerical evidence that the elementary
   excitations of this model transform under representation $\bs{\bar
     3}$ of SU(3) if the original spins of the model transform under
   rep.\ $\bs{3}$.  (2) a family of parent Hamiltonians for valence
   bond solids of SU(3) chains with spin reps.\ $\bs{6}$, $\bs{10}$,
   and $\bs{8}$ on each lattice site.  We argue that of these three
   models, only the latter two exhibit spinon confinement and a
   Haldane gap in the excitation spectrum.
 \end{abstract}



\maketitle

Beginning with the invention of the Bethe Ansatz in
1931~\cite{bethe31zp205} as a method to solve the $S=\frac{1}{2}$
Heisenberg chain with nearest neighbor interactions, a significant
share of the entire effort in condensed matter physics has been
devoted to the study of quantum spin chains.  Faddeev and
Takhtajan~\cite{faddeev-81pla375} discovered in 1981 that the
elementary excitations (now called spinons) of the spin-$1/2$
Heisenberg chain carry spin $1/2$ while the Hilbert space is spanned
by spin flips, which carry spin 1.  The fractional quantization of
spin in spin chains is comparable to the fractional quantization of
charge in quantized Hall liquids~\cite{laughlin83prl1395}.  In 1982,
Haldane~\cite{haldane83pla464} identified the O(3) nonlinear sigma
model as the effective low-energy field theory of SU(2) spin chains,
and argued that chains with integer spin possess a gap in the
excitation spectrum, while a topological term renders half-integer
spin chains gapless.

The general methods---the Bethe Ansatz and the use of effective field
theories including bosonization---are complemented by a number of
exactly solvable models, most prominently among them the
Majumdar--Ghosh (MG) Hamiltonian~\cite{majumdar-69jmp1399} for the
$S=\frac{1}{2}$ dimer chain, the AKLT model~\cite{affleck-87prl799} as
a paradigm of the gapped $S=1$ chain, and the Haldane--Shastry model
(HSM)~\cite{haldane88prl635, shastry88prl639, haldane91prl1529}.  In
the HSM the wave functions for the ground state and single-spinon
excitations are of a simple Jastrow form, elevating the conceptual
similarity to quantized Hall states to a formal equivalence.  One of
the unique features of the HSM is that the spinons are free in the
sense that they only interact through their half-Fermi
statistics~\cite{haldane91prl937,essler95prb13357,greiter-05prb224424}.
The HSM has been generalized from SU(2) to
SU($n$)~\cite{kawakami92prb1005,ha-92prb9359,bouwknegt-96npb345,
  schuricht-05epl987}.

For the MG and the AKLT model, only the ground states are known
exactly.  Nonetheless, these models have amply contributed to our
understanding of many aspects of spin chains, each of them through the
specific concepts captured in its ground
state~\cite{affleck89jpcm3047, okamoto-92pla433, eggert96prb9612,
  white-96prb9862, schollwock-96prb3304, kolezhuk-96prl5142,
  kolezhuk-02prb100401, normand-02prb104411}.

In the past, the motivation to study SU($n$) spin systems with $n>2$
has been mainly formal.  The Bethe Ansatz has been generalized to
multiple component systems by Sutherland~\cite{sutherland75prb3795},
and has been applied to the SU($n$)
HSM~\cite{kawakami92prb1005,ha-92prb9359}. 
The effective field theory description of Haldane yielding the
distinction between gapless half-integer spin chains and gapped
integer spin chains, however, cannot be directly generalized to
SU($n$) chains, as there is no direct equivalent of the CP$^1$
representation used in Haldane's analysis.

In recent years, ultracold atoms in optical lattices have provided a
framework for model realizations of various problems of condensed
matter physics, including the phase transition from a superfluid to a
Mott insulator~\cite{Jaksch-98prl3108,Greiner-02n39}, the fermionic
Hubbard model~\cite{Honerkamp-04prl170403,rapp0607138}, and SU(2) spin
chains~\cite{Duan-03prl090402,Garcia-04prl250405}.  
In particular, the Hamiltonians for spin lattice models may be
engineered with polar molecules stored in optical lattices, where the
spin is represented by a single-valence electron of a heteronuclear
molecule~\cite{micheli-06np341}.
Systems of ultracold atoms in optical lattices may further provide an
experimantal realization of SU(3) spin systems, and in particular
antiferromagnetic SU(3) spin chains, in due course.  A simple and
intriguing possibility is to manipulate an atomic system with total
angular momentum $F=\frac{3}{2}$ such that it simulates an SU(3)
spin.  For such atoms, one effectively suppresses
the occupation of one of the ``middle'' states, say the
$F^z=-\frac{1}{2}$ state, by shifting it to a higher energy while
keeping the other states approximately degenerate.  At sufficiently
low temperatures, one is left with three internal states
$F^z=-\frac{3}{2},+\frac{1}{2},+\frac{3}{2}$, which one identifies
with the colors ``blue'', ``red'', and ``green'' of an SU(3) spin.  In
leading order, the number of particles of each color is now conserved,
as required by the SU(3) symmetry.  If one places one of these atoms
at each site of an optical lattice and allows for a weak hopping, one
obtains an SU(3) antiferromagnet for sufficiently large on-site
repulsions $U$.

Motivated by both this prospect as well as the mathematical challenges
inherent to the problem, we propose several exact models for SU(3)
spin chains in this Letter.  The models are similar in spirit to the
MG or the AKLT model for SU(2), and consist of parent Hamiltonians and
their exact ground states.  

Consider a chain with $N$ lattice sites, where $N$ has to be divisible
by three, and periodic boundary conditions (PBCs). On each lattice
site we place an SU(3) spin which transforms under the fundamental
representation $\bs{3}=(1,0)$, \ie the spin can take the values (or
colors) blue (b), red (r), or green (g). (We label the representations
of SU(3) by their dimensions (the bold numbers) or their Dynkin
coordinates (a pair of non-negative integers)~\cite{Cornwell84vol2}.)
The trimer states are obtained by requiring the spins on each three
neighboring sites to form an SU(3) singlet $\bs{1}=(0,0)$, which we
call a trimer and sketch it by
\begin{picture}(38,8)(1,-2)\linethickness{0.8pt}
\put(6,0){\circle{4}}\put(8,0){\line(1,0){10}}\put(20,0){\circle{4}}
\put(22,0){\line(1,0){10}}\put(34,0){\circle{4}}
\end{picture}.
The three linearly independent trimer states 
are given by
\begin{equation}
\setlength{\unitlength}{1pt}
\begin{picture}(130,8)(4,-2)\linethickness{0.8pt}
\multiput(6,0)(14,0){9}{\circle{4}}
\multiput(8,0)(14,0){2}{\line(1,0){10}}
\multiput(50,0)(14,0){2}{\line(1,0){10}}
\multiput(92,0)(14,0){2}{\line(1,0){10}}
\end{picture}
\label{trimerstates.3}
\end{equation}
and two more states obtained by shifting this one by one or two
lattice sites, respectively.  Introducing operators
$c_{i\sigma}^{\dagger}$ which create a fermion of color $\sigma$
($\sigma=\b,\r,\g$) at lattice site $i$, the trimer states can be
written as
\begin{equation}
  \label{eq:trimer}
  \ket{\psi_{\text{trimer}}^{(\mu)}\!}=\hspace{-15pt}
  \prod_{\scriptstyle{i} \atop 
    \left(\scriptstyle{\frac{i-\mu}{3}\,{\rm integer}}\right)}
  \hspace{-5pt}\Bigl(
  \sum_{\scriptstyle{(\alpha,\beta,\gamma)}=
    \atop\scriptstyle{{\pi}(\b,\r,\g)}} 
  \hspace{-5pt}\hbox{sign}({\pi})\,
  c^{\dagger}_{i\,\alpha}\,c^{\dagger}_{i+1\,\beta}\,c^{\dagger}_{i+2\,\gamma}
  \Bigr)\!\vac\!,
\end{equation}
where $\mu=1,2,3$ labels the three degenerate ground states, $i$ runs
over the lattice sites subject to the constraint that
$\frac{i-\mu}{3}$ is integer, and the sum extends over all six
permutations $\pi$ of the three colors b, r, and g.

The SU(3) generators at each lattice site $i$ are defined as
\begin{equation}
J^a_i=\frac{1}{2}\,\sum_{\sigma,\sigma'=\b,\r,\g}\,
c_{i\sigma}^{\dagger} \lambda^a_{\sigma\sigma'}
c_{i\sigma'}^{\phantom{\dagger}}, \quad a=1,\ldots,8,
\label{eq:J_a}
\end{equation}
where the $\lambda^a$ are the Gell-Mann
matrices~\cite{Cornwell84vol2}.  The operators \eqref{eq:J_a} satisfy
the commutation relations
$\comm{J^a_{i}}{J^b_{j}}=\delta_{ij}\;f^{abc}J^c_{i}$,
$a,b,c=1,\ldots,8,$ with $f^{abc}$ the structure constants of SU(3).
We further introduce the total SU(3) spin of $\nu$ neighboring sites
$i,\ldots,i+\nu-1$,
\begin{equation}\label{eq:defJnu}
\bs{J}^{(\nu)}_i = \sum_{j=i}^{i+\nu-1}\,\bs{J}_{j},
\end{equation}
where $\bs{J}_i$ is the eight-dimensional vector formed by its
components \eqref{eq:J_a}.  The parent Hamiltonian for the trimer
states \eqref{eq:trimer} is given by
\begin{equation}
  \label{ham.trimer}
  H_{\textrm{trimer}}\, =\, \sum_{i=1}^N
  \left(\,\left(\bs{J}^{(4)}_i\right)^{\!4}\,
    -\,\frac{14}{3}\left(\bs{J}^{(4)}_i\right)^{\!2}\,+
    \,\frac{40}{9}\,\right).
\end{equation}
To verify this Hamiltonian, note that since the spins on the
individual sites transform under the fundamental representation
$\bs{3}$, the SU(3) content of four sites is
\begin{equation}\label{rep3power4}
\bs{3}\otimes\bs{3}\otimes\bs{3}\otimes\bs{3}\,=\,
3\cdot\bs{3}\,\oplus\,2\cdot\bs{\bar 6}\,\oplus\,3\cdot\bs{15}\,
\oplus\,\bs{15'},
\end{equation}
\ie we obtain representations $\bs{3}$, $\bs{\bar{6}}=(0,2)$, and the
two non-equivalent representations $\bs{15}=(2,1)$ and
$\bs{15'}=(4,0)$.  All these representations can be distinguished by
their eigenvalues of the quadratic Casimir operator~\cite{Cornwell84vol2}.

For the trimer states \eqref{trimerstates.3}, the situation
simplifies as we only have the two possibilities 
\begin{displaymath}
\begin{array}{rcl}
&
\begin{picture}(55,8)(0,-2)\linethickness{0.8pt}
\put(8,0){\line(1,0){10}}\put(6,0){\circle{4}}\put(20,0){\circle{4}}
\put(22,0){\line(1,0){10}}\put(34,0){\circle{4}}\put(48,0){\circle{4}}
\end{picture}
~&~\hat =~~~\bs{1}\,\otimes\,\bs{3}~=~\bs{3},\label{eq:one}\\[1mm]
&
\begin{picture}(55,8)(0,-2)\linethickness{0.8pt}
\put(8,0){\line(1,0){10}}\put(6,0){\circle{4}}
\put(20,0){\circle{4}}\put(36,0){\line(1,0){10}}\put(34,0){\circle{4}}
\put(48,0){\circle{4}}
\end{picture}~&~
\hat =~~~\bs{\bar 3} \,\otimes\,\bs{\bar 3}~=~\bs{3}\,\oplus\,\bs{\bar 6},
\end{array}
\end{displaymath}
(where $\bs{\bar{3}}=(0,1)$), \ie the total SU(3) spin on four
neighboring sites can only transform under representations $\bs{3}$ or
$\bs{\bar{6}}$.  The eigenvalues of the quadratic Casimir operator for
these representations are $\frac{4}{3}$ and $\frac{10}{3}$,
respectively.  The auxiliary operators
\begin{equation}
  \label{eq:auxop}
  H_i=\left(\Bigl(\bs{J}^{(4)}_i\Bigr)^2-\frac{4}{3}\right)
  \left(\Bigl(\bs{J}^{(4)}_i\Bigr)^2-\frac{10}{3}\right)
\end{equation} 
hence annihilate the trimer states for all values of $i$, while they
yield positive eigenvalues for $\bs{15}$ or $\bs{15'}$, \ie all other
states.  Summing $H_i$ over all lattice sites $i$ yields
\eqref{ham.trimer}.

\begin{figure}[t]
\centering
\includegraphics[scale=0.31,angle=270]{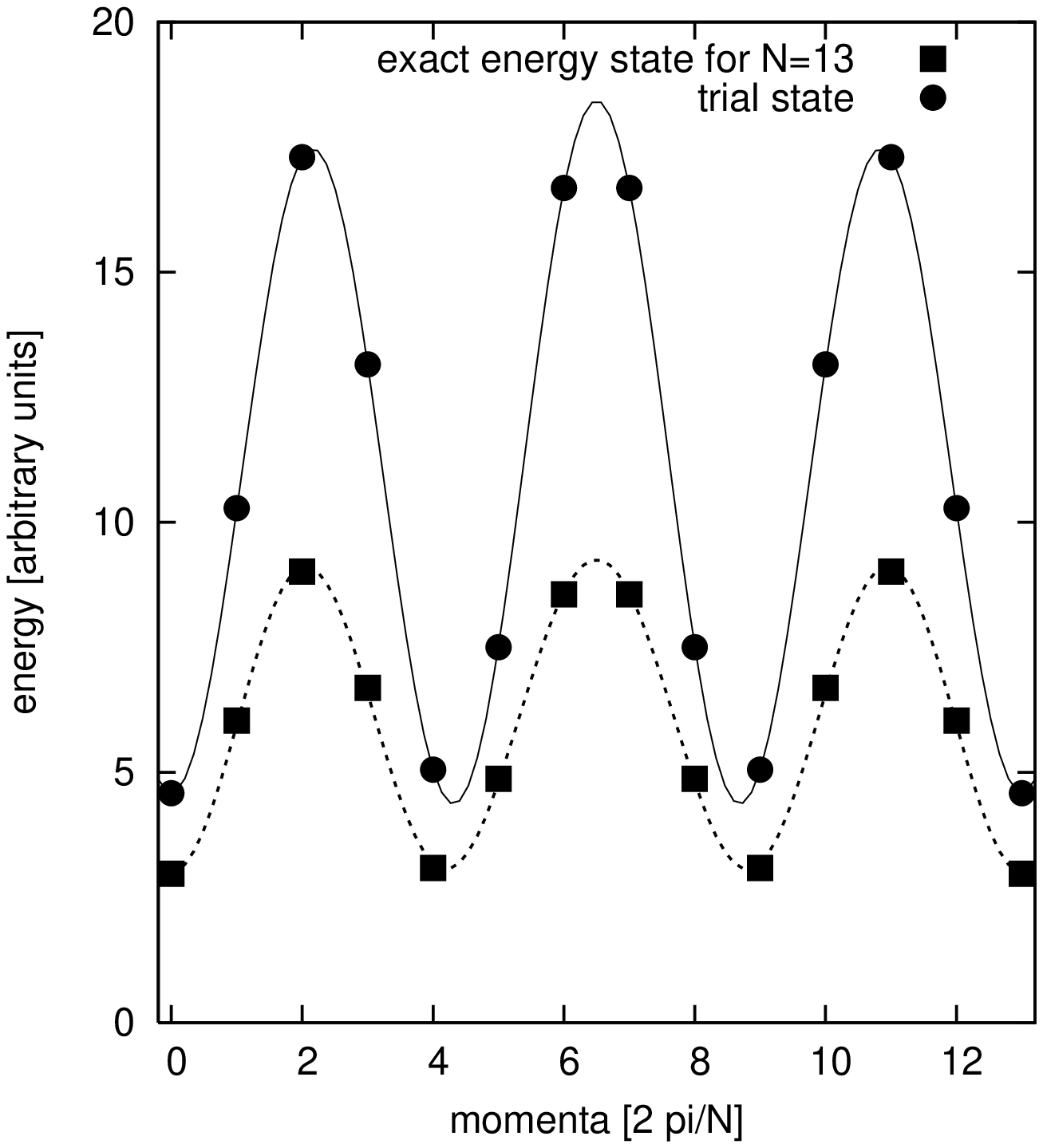}\hspace{-4mm}
\includegraphics[scale=0.31,angle=270]{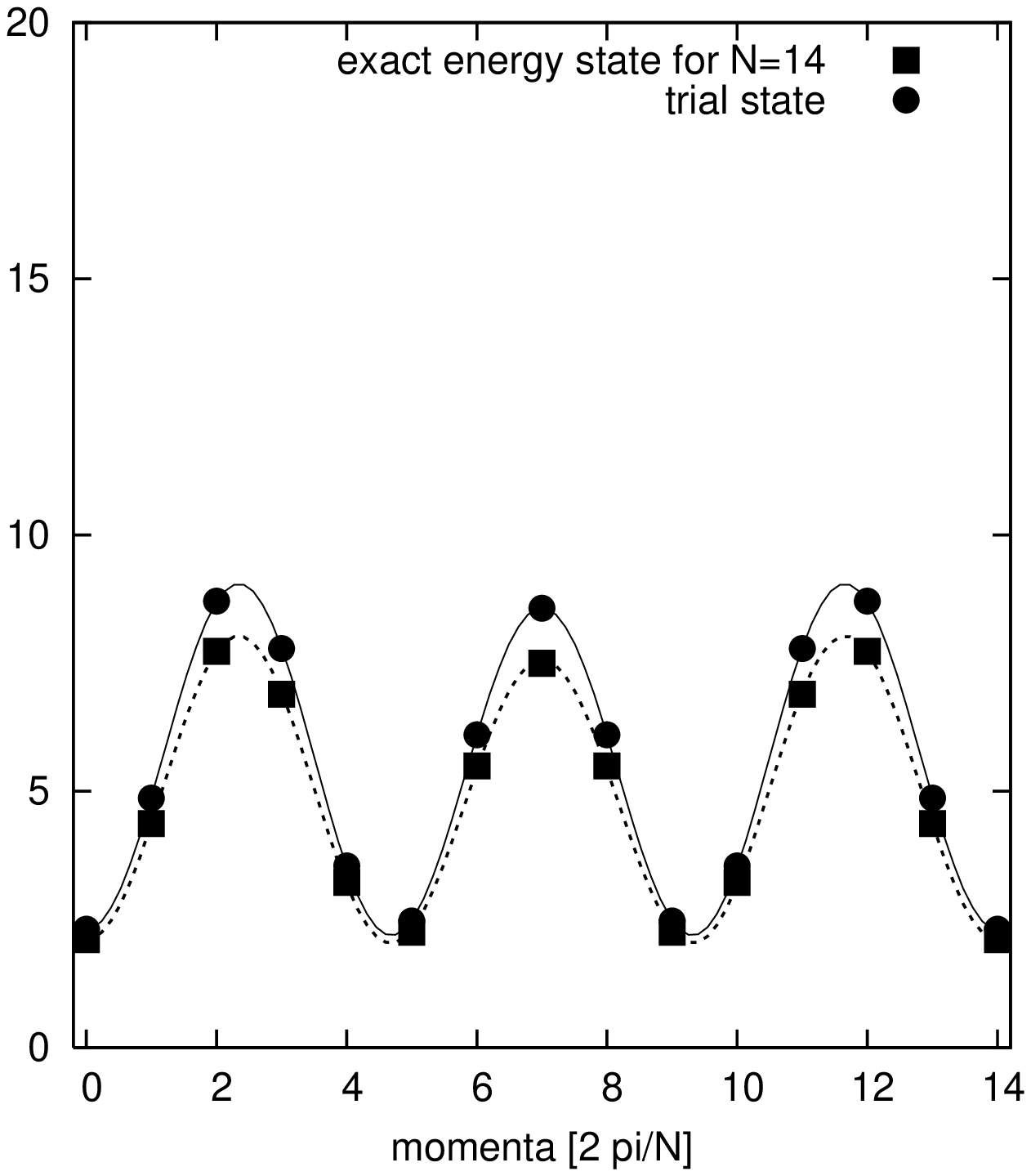}
\caption{Dispersions of the rep.\ $\bs{3}$ (left) and the rep.\
  $\bs{\bar{3}}$ trial states (right) in comparison to the exact
  excitation energies of \eqref{ham.trimer}.  The lines are a guide to
  the eye.}
\label{fig:exci}
\end{figure}

There are two different kinds of domain walls between the degenerate
ground states. The first kind consists of an individual SU(3) spin,
which transforms under representation $\bs{3}$; the second kind
consists of two antisymmetrically coupled spins on two neighboring
sites, and hence transforms under representation $\bs{\bar{3}}$.
Since they can decay into each other, only one of these domain walls
can constitute an approximate eigenstate of the trimer model.  We have
performed numerical studies on chains with $N=13$ and $N=14$, which
clearly indicate that the elementary excitations of the trimer chain
\eqref{ham.trimer} transform under $\bs{\bar{3}}$ (see
Fig.~\ref{fig:exci}).  This result appears to be a general feature of
rep.\ $\bs{3}$ SU(3) spin chains, as it was recently shown explicitly
to hold for the HSM as well~\cite{schuricht-05epl987}.  The elementary
excitations of the trimer chain are deconfined, meaning that the
energy of two localized representation $\bs{\bar{3}}$ domain walls or
colorons does not depend on the distance between them.

We now introduce a family of exactly soluble valence bond models for
SU(3) chains of various spin representations of the SU(3) spins at
each lattice site.  To formulate these models, we will use SU(3)
Schwinger bosons $b,b^\dagger$ (blue), $r,r^\dagger$ (red), and
$g,g^\dagger$ (green)~\cite{Auerbach94}, which are defined by
$\ket{\b}=c_{\b}^\dagger \vac=b^\dagger \vac$,
$\ket{\r}=c_{\r}^\dagger \vac=r^\dagger \vac$, and
$\ket{\g}=c_{\g}^\dagger \vac=g^\dagger \vac$, and satisfy
$\comm{b}{b^\dagger}=\comm{r}{r^\dagger}=\comm{g}{g^\dagger}=1$ while
all other commutators vanish. The Schwinger bosons can be used to
combine spins transforming under the fundamental representation
$\bs{3}=(1,0)$ symmetrically, and hence to construct representations
like $\bs{6}=(2,0)$ and $\bs{10}=(3,0)$.

The trimer states \eqref{eq:trimer} can be rewritten using SU(3) Schwinger
bosons as $\ket{\psi_\text{trimer}^{(\mu)}\!}=\Psi^{\mu}\!\left[
  b^{\dagger},r^{\dagger},g^{\dagger}\right]\,\vac$ with
\begin{equation}
\Psi^{\mu}\!\left[
    b^{\dagger},r^{\dagger},g^{\dagger}\right]=
  \!\!\!\prod_{\scriptstyle{i} \atop 
    \left(\scriptstyle{\frac{i-\mu}{3}\,{\rm integer}}\right)}
  \hspace{-5pt}\Bigl(
  \sum_{\scriptstyle{(\alpha,\beta,\gamma)}=\atop\scriptstyle{{\pi}(b,r,g)}} 
  \hspace{-5pt}\hbox{sign}({\pi})\,
  \alpha^{\dagger}_{i}\,\beta^{\dagger}_{i+1}\gamma^{\dagger}_{i+2}
  \Bigr).
\label{eq:trimerschw}
\end{equation}

We obtain a representations $\bs{6}$ VBS from two trimer states by
projecting the tensor product of two fundamental representations
$\bs{3}$ onto the symmetric subspace, \ie onto the $\bs{6}$ in the
tensor product $\bs{3}\,\otimes\,\bs{3}=\bs{\bar{3}}\,\oplus\,\bs{6}$.
Graphically, we illustrate this as follows:
\begin{equation}
\setlength{\unitlength}{1pt}
\begin{picture}(128,45)(4,-20)
\linethickness{0.8pt}
\multiput(0,10)(14,0){9}{\circle{4}}
\multiput(14,0)(14,0){9}{\circle{4}}
\thicklines
\put(2,10){\line(1,0){10}}
\put(16,10){\line(1,0){10}}
\put(44,10){\line(1,0){10}}
\put(58,10){\line(1,0){10}}
\put(86,10){\line(1,0){10}}
\put(100,10){\line(1,0){10}}
\put(16,0){\line(1,0){10}}
\put(30,0){\line(1,0){10}}
\put(58,0){\line(1,0){10}}
\put(72,0){\line(1,0){10}}
\put(100,0){\line(1,0){10}}
\put(114,0){\line(1,0){10}}
\thinlines
\put(65,-5){\framebox(10,20)}
\put(70,-15){\line(0,1){10}}
\put(70,-22){\makebox(1,1){\small projection onto rep.\ $\bs{6}=(2,0)$}}
\put(70,25){\makebox(1,1){\small one site}}
\end{picture}
\label{fig:6state}
\end{equation}
This construction yields three linearly independent $\bs{6}$ VBS
states, which are readily written out using \eqref{eq:trimerschw},
\begin{equation}
\label{eq:6VBS}
\ket{\psi_{\bs{6}\, \text{VBS}}^{(\mu)}}=
\Psi^{\mu}\!\left[b^{\dagger},r^{\dagger},g^{\dagger}\right]\cdot
\Psi^{\mu+1}\!\left[b^{\dagger},r^{\dagger},g^{\dagger}\right]\,
\vac
\end{equation}
for $\mu=1$, 2, or 3.  These states are zero-energy ground states of
the parent Hamiltonian $H_{\bs{6}\,\text{VBS}}=\sum_{i=1}^N H_i$
with 
\begin{equation}
  \label{eq:hi6}
  H_i=\left(\!\left(\!\bs{J}^{(4)}_i\!\right)^2 \! - \frac{4}{3}\right)
  \!\left(\!\left(\!\bs{J}^{(4)}_i\!\right)^2 \! - \frac{10}{3}\right)
  \!\left(\!\left(\!\bs{J}^{(4)}_i\!\right)^2 \! - \frac{16}{3}\right)
\end{equation}
(see \cite{Greiter-inpreparation}).
Note that the operators $J_i^a$, $a=1,\ldots,8$, are now given by
$6\times 6$ matrices, as the Gell-Mann matrices only provide the
generators \eqref{eq:J_a} of the fundamental representation $\bs{3}$.
As in the trimer model, two distinct types of domain walls exist,
which transform according to reps.\ $\bs{3}$ and $\bs{\bar{3}}$.  
Since both excitations are merely domain walls between different
ground states, there is no confinement between them.

Let us now turn to the $\bs{10}$ VBS chain.  By combining the three
different trimer states \eqref{eq:trimerschw} symmetrically,
\begin{equation}
\ket{\psi_{\bs{10}\,\text{VBS}}}\!\!=\!\!
\Psi^1\!\left[b^{\dagger}\!,r^{\dagger}\!,g^{\dagger}\right]
\!\cdot\!\Psi^2\!\left[b^{\dagger}\!,r^{\dagger}\!,g^{\dagger}\right]
\!\cdot\!\Psi^3\!\left[b^{\dagger}\!,r^{\dagger}\!,g^{\dagger}\right]\vac, 
\label{state.10VBS}
\end{equation}
we automatically project out the rep.\ $\bs{10}$ in the tensor product
$\bs{3}\otimes\bs{3}\otimes\bs{3}=\bs{1}\oplus2\!\cdot\bs{8}\oplus\bs{10}$
generated on each lattice site by the three trimer chains.  This
construction yields a unique state:
\begin{equation}
\setlength{\unitlength}{1pt}
\begin{picture}(180,55)(4,-4)
\linethickness{0.8pt}
\multiput(0,35)(14,0){12}{\circle{4}}
\multiput(14,25)(14,0){12}{\circle{4}}
\multiput(28,15)(14,0){12}{\circle{4}}
\thicklines
\put(2,35){\line(1,0){10}}
\put(16,35){\line(1,0){10}}
\put(44,35){\line(1,0){10}}
\put(58,35){\line(1,0){10}}
\put(86,35){\line(1,0){10}}
\put(100,35){\line(1,0){10}}
\put(16,25){\line(1,0){10}}
\put(30,25){\line(1,0){10}}
\put(58,25){\line(1,0){10}}
\put(72,25){\line(1,0){10}}
\put(100,25){\line(1,0){10}}
\put(114,25){\line(1,0){10}}
\put(30,15){\line(1,0){10}}
\put(44,15){\line(1,0){10}}
\put(72,15){\line(1,0){10}}
\put(86,15){\line(1,0){10}}
\put(114,15){\line(1,0){10}}
\put(128,15){\line(1,0){10}}
\put(128,35){\line(1,0){10}}
\put(142,35){\line(1,0){10}}
\put(142,25){\line(1,0){10}}
\put(156,25){\line(1,0){10}}
\put(156,15){\line(1,0){10}}
\put(170,15){\line(1,0){10}}
\thinlines
\put(51,10){\framebox(10,30){}}
\put(56,2){\line(0,1){8}}
\put(56,-5){\makebox(1,1){\small projection onto $\bs{10}=(3,0)$}}
\put(56,48){\makebox(1,1){\small one site}}
\put(121,10){\dashbox{2}(24,30)}
\end{picture}
\label{fig:10state}
\end{equation}
The parent Hamiltonian acts on pairs of neighboring sites. It is
constructed by noting that the only representations which are included
in both $\bs{10}\otimes\bs{10}$ and \linebreak
$\bs{\bar{3}}\otimes\bs{\bar{3}}\otimes\bs{3}\otimes\bs{3}$ (which is
the representation content of the total spin of two neighboring sites
of the VBS state as indicated by the dashed box above) are the reps.\
$\bs{\overline{10}}=(0,3)$ and $\bs{27}=(2,2)$. With the eigenvalues
of the Casimir operator, which are $6$ and $8$, respectively, we
obtain the parent Hamiltonian for \eqref{state.10VBS}
\begin{equation}
  \label{ham.10VBS}
  H_{\bs{10}\,\text{VBS}}
  =\sum_{i=1}^N\,\left(\bigl(\bs{J}_i\bs{J}_{i+1}\bigr)^2
    +5\,\bs{J}_i\bs{J}_{i+1}+6\right).
\end{equation}

The Hamiltonian \eqref{ham.10VBS} provides the equivalent of the AKLT
model~\cite{affleck-87prl799}, whose unique ground state is
constructed from dimer states by projection onto spin 1, for SU(3)
spin chains.  

Since the $\bs{10}$ VBS state \eqref{state.10VBS} is unique, domain
walls connecting different ground states do not exist.  We hence
expect the coloron and anti-coloron excitations to be confined in
pairs, as illustrated below.  The state between the excitations is no
longer annihilated by \eqref{ham.10VBS}, as there are pairs of
neighboring sites containing representations different from
$\bs{\overline{10}}$ and $\bs{27}$, as indicated by the dotted box
below.  As the number of such pairs increases linearly with the
distance between the excitation, the confinement potential depends
linearly on this distance.
\begin{equation}
\setlength{\unitlength}{1pt}
\begin{picture}(210,66)(0,2)
\linethickness{0.8pt}
\multiput(0,45)(14,0){15}{\circle{4}}
\multiput(42,45)(14,0){2}{\circle*{4}}
\multiput(154,45)(14,0){1}{\circle*{4}}
\multiput(14,35)(14,0){15}{\circle{4}}
\multiput(28,25)(14,0){15}{\circle{4}}
\multiput(182,25)(14,0){3}{\circle{4}}
\thicklines
\put(2,45){\line(1,0){10}}\put(16,45){\line(1,0){10}}
\put(44,45){\line(1,0){10}}\put(72,45){\line(1,0){10}}
\put(86,45){\line(1,0){10}}\put(114,45){\line(1,0){10}}
\put(128,45){\line(1,0){10}}\put(170,45){\line(1,0){10}}
\put(184,45){\line(1,0){10}}
\put(16,35){\line(1,0){10}}\put(30,35){\line(1,0){10}}
\put(58,35){\line(1,0){10}}\put(72,35){\line(1,0){10}}
\put(100,35){\line(1,0){10}}\put(114,35){\line(1,0){10}}
\put(142,35){\line(1,0){10}}\put(156,35){\line(1,0){10}}
\put(184,35){\line(1,0){10}}\put(198,35){\line(1,0){10}}
\put(30,25){\line(1,0){10}}\put(44,25){\line(1,0){10}}
\put(72,25){\line(1,0){10}}\put(86,25){\line(1,0){10}}
\put(114,25){\line(1,0){10}}\put(128,25){\line(1,0){10}}
\put(156,25){\line(1,0){10}}\put(170,25){\line(1,0){10}}
\put(198,25){\line(1,0){10}}\put(212,25){\line(1,0){10}}
\thinlines
\put(105,11){\vector(-1,0){49}}
\put(105,11){\vector(1,0){49}}
\put(105,2){\makebox(0,0){\small energy cost $\propto$ distance}}
\put(49,56){\makebox(0,0){$\bs{\bar 3}$}}
\put(154,56){\makebox(0,0){$\bs{3}$}}
\put(49,67){\makebox(0,0){\small coloron}}
\put(154,67){\makebox(0,0){\small anti-coloron}}
\put(93,20){\dashbox{1}(24,30)}
\end{picture}\label{10VBSstate.exc}
\end{equation}
The confinement force between the pair induces a linear oscillator
potential for the relative motion of the constituents.  The zero-point
energy of this oscillator gives rise to a Haldane-type energy gap
(see~\cite{greiter02prb134443} for a similar discussion in the two-leg
Heisenberg ladder).  We expect this gap to be a generic feature of
rep.\ $\bs{10}$ spin chains with short-range antiferromagnetic
interactions.

Finally, we construct a representation $\bs{8}$ VBS state, where
$\bs{8}=(1,1)$ is the adjoint representation of SU(3). Consider first a chain
with alternating reps.\ $\bs{3}$ and $\bs{\bar{3}}$ on neighboring
sites, which we combine into singlets.  This can be done in two ways,
yielding the two states
\begin{equation}
\setlength{\unitlength}{1pt}
\begin{picture}(210,14)(0,-10)
\linethickness{0.8pt}
\multiput(5,0)(30,0){3}{\circle{4}}
\multiput(20,0)(30,0){3}{\circle{6}}
\multiput(138,0)(30,0){3}{\circle{4}}
\multiput(153,0)(30,0){3}{\circle{6}}
\thicklines
\multiput(7,0)(30,0){3}{\line(1,0){10}}
\multiput(156,0)(30,0){2}{\line(1,0){10}}
\put(136,0){\line(-1,0){4}}
\put(216,0){\line(1,0){3}}
\thinlines
\put(109,0){\makebox(0,0){and}}
\put(224,-3){\makebox(0,0){.}}
\put(37.7,-10){\makebox(0,0){\small $\bs{3}\phantom{\bs{\bar{3}}}$}}
\put(50,-10){\makebox(0,0){\small $\bs{\bar{3}}$}}
\end{picture}
\nonumber
\end{equation}
We then combine one $\bs{3}$-$\bs{\bar{3}}$ state with the one
shifted by one lattice spacing.  This yields representations
$\bs{3}\otimes\bs{\bar{3}}=\bs{1}\oplus\bs{8}$ at each site.  The
$\bs{8}$ VBS state is obtained by projecting onto the adjoint 
representations $\bs{8}$.  Corresponding to the two
$\bs{3}$-$\bs{\bar{3}}$ states illustrated above, we obtain two
linearly independent $\bs{8}$ VBS states, $\Psi^\text{L}$ and
$\Psi^\text{R}$, which may be visualized as
\begin{equation}
\setlength{\unitlength}{1pt}
\begin{picture}(210,50)(0,-25)
\linethickness{0.8pt}
\multiput(5,10)(30,0){3}{\circle{4}}
\multiput(20,10)(30,0){3}{\circle{6}}
\multiput(138,10)(30,0){3}{\circle{4}}
\multiput(153,10)(30,0){3}{\circle{6}}
\multiput(5,0)(30,0){3}{\circle{6}}
\multiput(20,0)(30,0){3}{\circle{4}}
\multiput(138,0)(30,0){3}{\circle{6}}
\multiput(153,0)(30,0){3}{\circle{4}}
\thicklines
\multiput(7,10)(30,0){3}{\line(1,0){10}}
\multiput(156,10)(30,0){2}{\line(1,0){10}}
\multiput(22,0)(30,0){2}{\line(1,0){10}}
\multiput(141,0)(30,0){3}{\line(1,0){10}}
\put(136,10){\line(-1,0){4}}
\put(216,10){\line(1,0){3}}
\put(2,0){\line(-1,0){3}}
\put(82,0){\line(1,0){4}}
\thinlines
\put(109,5){\makebox(0,0){and}}
\put(224,2){\makebox(0,0){.}}
\put(44,-6){\framebox(12,22)}
\put(50,-6){\line(0,-1){9}}
\put(50,-22){\makebox(1,1){\small projection onto $\bs{8}=(1,1)$}}
\put(50,25){\makebox(1,1){\small one site}}
\end{picture}
\label{fig:8VBS}
\end{equation}
These states transform into each other under parity or color
conjugation (interchange of $\bs{3}$ and $\bs{\bar{3}}$).  The
corresponding states may be formulated as a matrix
product~\cite{Greiter-inpreparation}.

The parent Hamiltonian for these states is constructed along the same lines as
above, yielding 
\begin{equation}
\label{ham.8VBS}
H_{\bs{8}\,\text{VBS}}=\sum_{i=1}^N\left(\bigl(\bs{J}_i\bs{J}_{i+1}\bigr)^2+
  \frac{9}{2}\,\bs{J}_i\bs{J}_{i+1} + \frac{9}{2}\right).
\end{equation}
$\Psi^\text{L}$ and $\Psi^\text{R}$ are the only (zero-energy) ground 
states of \eqref{ham.8VBS} for $N\ge 3$.

The low-energy excitations of the $\bs{8}$ VBS model are given by
coloron--anti-coloron bound states:
\begin{equation}
\setlength{\unitlength}{1pt}
\begin{picture}(175,50)(0,-18)
\linethickness{0.8pt}
\multiput(5,10)(30,0){6}{\circle{4}}
\multiput(20,10)(30,0){6}{\circle{6}}
\multiput(35,0)(30,0){5}{\circle{6}}
\multiput(20,0)(30,0){5}{\circle{4}}
\multiput(125,0)(30,0){1}{\circle*{6}}
\multiput(50,0)(30,0){1}{\circle*{4}}
\thicklines
\multiput(7,10)(30,0){6}{\line(1,0){10}}
\multiput(142,0)(30,0){1}{\line(1,0){10}}
\multiput(22,0)(30,0){1}{\line(1,0){10}}
\multiput(68,0)(30,0){2}{\line(1,0){10}}
\thinlines
\put(22,-15){\makebox(0,0){\small $\Psi^\text{L}$}}
\put(157,-15){\makebox(0,0){\small $\Psi^\text{L}$}}
\put(50,23){\makebox(0,0){\small $\bs{3}$}}
\put(125,23){\makebox(0,0){\small $\bs{\bar{3}}$}}
\put(50,34){\makebox(0,0){\small anti-coloron}}
\put(125,34){\makebox(0,0){\small coloron}}
\put(87.5,-14){\vector(-1,0){37.5}}
\put(87.5,-14){\vector(1,0){37.5}}
\put(87.5,-23){\makebox(0,0){\small energy cost $\propto$ distance}}
\end{picture}
\label{fig:8VBS33bar}
\end{equation}
We find a linear confinement potential between the excitations, and
hence a Haldane-type gap in the spectrum.  Numerical studies on
a chain with $N=8$ sites provide evidence in support of this
conclusion~\cite{Greiter-inpreparation}.  In addition to the bound
state \eqref{fig:8VBS33bar}, the model allows for domain wall between
the two ground states $\Psi^\mathrm{L}$ and $\Psi^\mathrm{R}$.  They
consist of bound states of either two colorons or two anti-colorons,
which are confined through the same mechanisms as the
coloron--anti-coloron bound state \eqref{fig:8VBS33bar}, as one
may easily infer from a cartoon similar to the one above.  We hence
expect a Haldane gap for each individual domain wall as well.

The results regarding confinement and deconfinement of the excitations
of the $\bs{6}$, $\bs{10}$, and $\bs{8}$ VBS presented here are
consistent with a rigorous theorem by Affleck and
Lieb~\cite{affleck-86lmp57} on the existence of energy gaps in the
spectrum of SU($n$) nearest-neighbor Heisenberg spin chains.  

In conclusion, we have formulated several exact models of SU(3) spin
chains.  We first introduced a trimer model and presented evidence that
the elementary excitations of the model transform under the SU(3)
representations conjugate to the representation of the original spin
on the chain. We further introduced three SU(3) valence bond solid
chains with representation $\bs{6}$, $\bs{10}$, and $\bs{8}$,
respectively, on each lattice site.  The elementary excitations of the
$\bs{10}$ and the $\bs{8}$ valence bond solid chain were found to be
confined.  

We would like to thank Peter Zoller, Ronny Thomale, and Peter W\"olfle
for valuable discussions of various aspects of this work.  SR was
supported by a Ph.D.  scholarship of the Cusanuswerk, and DS by the
German Research Foundation (DFG) through GK 284 and the Center for
Functional Nanostructures Karlsruhe.


\end{document}